\documentstyle[12pt,epsf,equation]{article}
\begin{document}
\pagestyle{empty}
\def\to{\rightarrow}
\def\rs{\mbox{$\sqrt{s}$}}
\def\pt{\mbox{$p_T$}}
\def\ccbar{\mbox{$c \bar c$}}
\def\f2gam{\mbox{$F_2^\gamma$}}
\def\gstarga{\mbox{$\gamma^* \gamma$}}
\def\gamgam{\mbox{$\gamma \gamma$}}
\def\eplem{\mbox{$e^+e^-$}}
\def\glgam{\mbox{$g^\gamma (x,Q^2)$}}
\def\ep{\mbox{$ep$}}
\def\gamp{\mbox{$\gamma p$}}
\def\qgam{\mbox{$q^\gamma (x,Q^2)$}}
\def\fpigam{\mbox{$f_{P_{i}|\gamma}$}}
\def\fgame{\mbox{$f_{\gamma |e} $}}
\def\fqgam{\mbox{$f_{q|\gamma}$}}
\def\fggam{\mbox{$f_{g|\gamma}$}}
\def\pcsq{\mbox{$P_c^2$}}
\def\pc{\mbox{$P_c$}}
\def\fpie{\mbox{$f_{P_{i}|e} $}}
\def\fie{\mbox{$f_{i|e} $}}
\def\fqe{\mbox{$f_{q|e} $}}
\def\fge{\mbox{$f_{g|e} $}}
\def\fcgam{\mbox{$f_{c|\gamma}$}}
\def\fp2p{\mbox{$f_{P_2|p}$}}
\def\fgp{\mbox{$f_{g/p}$}}
\def\qsq{\mbox{$Q^2$}}
\def\psqmax{\mbox{$P^2_{\rm {max}}$}}
\def\psqmin{\mbox{$P^2_{\rm {min}}$}}
\def\psq{\mbox{$P^2$}}
\def\ftilde{\mbox{$\tilde f$}}
\def\frest{\mbox{$f^{\rm rest}$}}
\def\ben{\begin{subequations}}
\def\een{\end{subequations}}
\def\beq{\begin{eqalignno}}
\def\eeq{\end{eqalignno}}
\def\be{\begin{equation}}
\def\ee{\end{equation}}
\def\bea{\begin{eqnarray}}
\def\eea{\end{eqnarray}}
\hfill{LNF--95/027 (P)}
\vskip 0.1 cm
\hfill{MAD--PH--891}
\vskip 0.1 cm
\hfill{May 1995}
\vskip 0.5 cm
\begin{center}
{\large \bf Modification of the equivalent photon approximation (EPA)
 for `resolved' photon processes.\footnote{Talk presented by R.M. Godbole
at Photon'95, Sheffield, UK, April 8-13,1995.}}
\vglue 0.5 cm
Manuel Drees\footnote{Heisenberg Fellow}\\
{\sl
University of Wisconsin, Dept. of Physics, 1150 University Avenue,
Madison, WI 53706, U.S.A.}
\vglue 0.25 cm
and \\
\vglue 0.25 cm
Rohini M. Godbole\footnote{Permanent address:
Physics Department, University of Bombay, Vidyanagari, Bombay - 400 098,
India }\\
{\sl INFN--Laboratori Nazionali di Frascati, P.O.Box 13, I-00044,
Frascati (Roma), Italy}
\end{center}
\vglue 0.5cm
\begin{abstract}
\noindent
We propose a modification of the equivalent photon approximation (EPA) for
processes  which involve the parton content of the photon, to take into
account the suppression of the photonic parton fluxes due to the virtuality of
the photon. We present simple, physically motivated ans\"atze to model this
suppression and show that even though the parton content of the electron no
longer factorizes into an electron flux function and photon structure
function, it is still possible to express it as a single integral. We also
show that for the TRISTAN experiments its effect  can be numerically of the
same size  as that of the  NLO corrections. Further, we discuss a possible
measurement at HERA, which can provide an experimental handle on the effect we
model through our ans\"atze.
\end{abstract}
\par \vskip 0.1in \noindent
\clearpage
\setcounter{page}{1}
\pagestyle{plain}

Studies of jet production in  \gamgam\ (\eplem) collisions at TRISTAN
\cite{1} and LEP \cite{2} and in \gamp\ (\ep) collisions at HERA
\cite{3} have yielded clear evidence for hard scattering from partons in the
photon or the so called `resolved' processes \cite{4}. Having confirmed
the existence of these contributions to jet production, the next step is to use
them to get further information \cite{5} about the parton content of the
photon,
especially \glgam , about which very little direct information
is available so far. To that end one needs to address the question of
uncertainties in the theoretical predictions of these cross-sections.
This implies that the approximations made in the calculations need to be
improved. In this short note we study the issue of improvement of one of
these approximations.

Theoretical calculations for the \eplem\ and \ep\ processes are usually
done in the framework of the Weizs\"acker--Williams (WW) approximation also
alternatively called the equivalent photon approximation (EPA) \cite{6,7}. In
this approximation  the cross--section for a process $e + X \to e + X'$
where a $\gamma$ is exchanged in the $t/u$ channel, is given in
terms of the cross--section for the process $\gamma + X \to X'$ (for an {\it
on-shell} $\gamma$)  and the flux factor $\fgame (z)$ for a photon to carry
energy fraction $z$ of the $e$. For example, the cross--section for jet
production in $ep$ collisions (with $X = p$ and $X' = {\rm {jets}}$)  can be
written as
\be
{d \sigma \over dp_T}(ep \to {\rm {jets}}) =  \int_{z_{\rm {min}}}^1
\fgame (z)\;
{d \sigma \over dp_T}(\gamma p \to {\rm {jets}})|_{\hat {s} = s \cdot z}
  \; dz \; ,
\label{epa}
\ee
where $s$ is the squared centre--of--mass energy of the $ep$ system, and
$z_{min}$ is the minimum energy fraction that the $\gamma$ has to carry in
order for the process to be kinematically possible.
This approximation is valid only if (i) the major contribution to the
cross--section for the full process comes from the region where the virtuality
of the photons \psq\ is small compared to  \qsq ,
where \qsq\ is the typical scale characterising the process (say $p_T^2$ in the
present case), and (ii) the contribution of longitudinal photons to the
cross--section is very small. In this approximation, neglecting any $P^2$
dependence that the cross--section $\gamma + p \to {\rm {jets}}$ may have,
 the flux factor \fgame\ is given by  \cite{6},
\ben \label{epa1}
\beq
\fgame (z) & =\int_{\psqmin}^{\psqmax} \ftilde (z)\;{dP^2 \over P^2}
\;\; - \;\;
{\alpha \over 2 \pi} m_e^2 \cdot z \left[ {1 \over \psqmax} - {1 \over
\psqmin}\right] \label{epa1a} \\
 & = \ln \left( {\psqmax \over \psqmin}\right)\;  \ftilde (z)\; - \; \frest
(z).
\label{epa1b}
\eeq
\een
Here \frest\ is simply the second term in the first line  of Eq. (\ref{epa1})
 and $\ftilde (z)$ is just the usual WW splitting function.
In Eq. (\ref{epa1}) the  kinematic values for the limits \psqmax\  , \psqmin\
on virtuality  {\it  viz.} $P^2_{\rm {max, kin}}$ , $P^2_{\rm {min, kin}}$
are determined by the (anti)tagging conditions in a particular experiment.
It is also clear from the conditions of the validity of the approximations
that the upper limit on the virtuality has to be less than ${\cal O}(\qsq )$.
Taking this into account one has for \psqmax\ and \psqmin\ in Eq. (\ref{epa1})
\be
P^2_{\rm {max}}  = \min \; (P^2_{\rm {max,kin}} , \kappa Q^2 ) ; \;\;\;
P^2_{\rm {min}}  =  P^2_{\rm {min,kin}}.
\label{finlim}
\ee
$\kappa$ is a number $ \sim {\cal O} (1) $ whose proper value can
 be determined \cite{6,7} in processes where
the $\gamma$ directly participates in the hard process, by comparing the
results of the exact calculation with that given by Eq. (\ref{epa}).
For `resolved' processes  such an estimation of $\kappa$ is not available.
However, since the contribution from the large \psq\ region to these processes
is suppressed very strongly the exact value of $\kappa$ is not important,
except for realising that if $\psq > \qsq$ the concept of partons `in' the
photon will not make much sense at all. We therefore set $\kappa = 1$.

The important point is that for `resolved' processes, the approximation of
neglecting the \psq\ dependence of  ${d \sigma \over d p_T} (\gamma + p \to
{\rm jets}) $ is not quite correct as it involves the parton content of the
photon which can have an additional \psq\ dependence which is not simply of the
type $\psq / \qsq$ as there is an additional scale $\Lambda^2$ in this
case. The `resolved' contribution to this process is normally given by
\be
{d \sigma \over dp_T} (e p \to {\rm jets})  =  \sum_{P_i} \int dy \;
dx_2 \; \fpie (y,\qsq)\; \fp2p (x_2, \qsq)
{d \hat{\sigma} \over dp_T} (P_1 + P_2 \to P_3 + P_4),
\label{res}
\ee
where
\be
\fpie (y,\qsq) = \int_y^1 {dx \over x} \fgame \left( {y \over x} \right)
\fpigam (x,\qsq),
\label{iel}
\ee
and \fgame\ is given by Eq. (\ref{epa1b}). To take into account the effect of
the virtuality of the photon on its parton content and the resultant
modification of the EPA, we generalise Eq. (\ref{iel})
(using Eq. (\ref{epa1a})) as,
\be
\fpie (y,\qsq) =  \int_y^1 {dx \over x}
\biggl[ \ftilde \biggl( {y \over x} \biggr) \int_{\psqmin}^{\psqmax}
{d\psq \over \psq} \fpigam (x,\qsq,\psq)
+ \frest \biggl( {y \over x} \biggr) \fpigam (x,\qsq,0) \biggr].
\label{virtie}
\ee
Since \frest\ is non--negligible only for $\psqmin \ll \Lambda^2$, in the
second term we can drop the \psq\ dependence of the parton density in the
photon and they are thus the same as those measured in the DIS experiments.
For simplicity we will omit the second term in our subsequent expressions,
although it will be taken into account in our numerical results.

The real question now concerns the \psq\ dependence of $\fpigam (x,\qsq,\psq)$.
For $\qsq \gg \psq \gg \Lambda^2$ this can be computed rigorously in
perturbative QCD (pQCD) \cite{8,9}. These calculations \cite{9}
tell us that  the parton densities in a virtual $\gamma$ simply approach the
QPM predictions as $\psq \to \qsq$, while the \psq\ dependence vanishes
altogether if $\psq \ll \Lambda^2$. For the region $\psq \simeq \Lambda ^2$,
however, we do not have any information from these calculations. These
calculations \cite{9} also show that the gluon density in the
virtual photon is more suppressed than the quark densities. This is also
reasonable as the gluon in the photon arises from radiation off a quark,
so the further away the photon (hence the quark) from the mass--shell,
the more suppressed will be the gluon densities. Thus we have for large \qsq\
\ben
\label{virtqcd}
\beq
f_{q_i|\gamma} (x,\qsq,\psq) & = \qgam    \;\;\;\;\;\;\;\;\;\;\;\;\;
\psq \ll \Lambda^2 \nonumber \\
& = c_q^{\rm QPM} \; \ln {\qsq \over \psq} \;\;\;\;\;\;\;\;\;\;
\psq  \gg \Lambda^2
\label{virtqcdq}\\
f_{g|\gamma} (x,\qsq,\psq) & = \glgam      \;\;\;\;\;\;\;\;\;\;\;\;\;
\psq \ll \Lambda^2 \nonumber \\
& \propto  \ln^2 \; {\qsq \over \psq}        \;\;\;\;\;\;\;\;\;\;
\psq  \gg \Lambda^2
\label{virtqcdg}.
\eeq
\een
Here \qgam\ and \glgam\ are just the quark and gluon densities in an on--shell
$\gamma$. Using this information as guideline we can propose ans\"atze for
$f_{P_i|\gamma} (x,\qsq,\psq)$ which will interpolate smoothly between the two
abovementioned behaviours.  Because of the different \psq\ dependence of the
quark and gluon densities in the photon at large \psq\ we have to treat the
two separately. In modelling $f_{P_i|\gamma} (x,\qsq,\psq)$ for the
intermediate region $\psq \simeq \Lambda^2$ we do not attempt to produce
exactly either the \psq\ dependence at a fixed $x$ or the $x$ dependence at a
fixed \psq . We try instead to model the overall effect in terms of a single
parameter $P_c$. The simplest ansatz is
\beq
\fqgam^{(1)}(x,\qsq,\psq) &= \qgam ,\;\;\;\;\; \psq \leq \pcsq \nonumber \\
             &= c_q (x,\qsq) \;  \ln {\qsq \over \psq}, \;\;\; \psq \geq \pcsq
{}.
\label{qansatz1}
\eeq
Here $P_c$ is a free parameter of typical hadronic scale, and
continuity at $\psq = \pcsq$ determines $c_q(x,\qsq)$:
\be
c_q(x,\qsq) = {\qgam \over \ln(\qsq /\pcsq)}.
\label{cq}
\ee
However, since this ansatz is motivated by the pQCD result, one can also
alternatively have a parameter free ansatz where one uses instead the QPM
prediction for $c_q$:
\be
c_q(x,\qsq) =
c_q^{QPM}(x) = 3 {\alpha \over 2 \pi} e_q^2 \; [x^2 + (1-x)^2],
\label{cqqpm}
\ee
in Eq. (\ref{qansatz1}), and  then solve Eq. (\ref{cq}) for $P_c$.
The $P_c$ so calculated will in general depend on $x$ and \qsq. The fact
that the values of $P_c$ so obtained are typically of hadronic scale gives us
confidence in our ansatz. The simple ansatz of Eq. (\ref{qansatz1}), however,
has a kink when plotted as a function of $\ln \psq$. We can smooth this out by
writing
\be
\fqgam^{(2)}(x,\qsq,\psq) = \qgam  {\ln {\qsq +\psq \over \psq + \pcsq} \over
                 \ln(1 + \qsq/\pcsq)}.
\label{qansatz2}
\ee
This ansatz now interpolates smoothly between the two limits $\psq \to 0 $ and
$\psq \to \qsq $. The nice feature of these two ans\"atze is that when
we put these back in Eq. (\ref{virtie}), we can once again write  \fqe\ as
\be
\fqe (y) = \int_y^1 {dx \over x} \ftilde \left( {y \over x} \right) \qgam
{\cal H}_q(\qsq,\psq,\pcsq,\psqmax,\psqmin),
\label{effeie}
\ee
where ${\cal H}_q$ is an analytic function \cite{10}. In the case of ansatz of
Eq. (\ref{qansatz2}) this is possible only as an approximation, but the
approximate analytical expression given in Eq. (17) of Ref. \cite{10} is
accurate to better than  2\%. Thus even though we have lost the exact
factorization of the density of partons in an electron into a photon
flux factor and partonic densities in the photon, the expressions obtained by
us are no more complicated than usual to use in a numerical
calculation since we can still write \fpie\ as  a single integral
as in Eq. (\ref{iel}) before.

The  ans\"atze for the gluon density $\fggam (x,\qsq,\psq) $ will have to take
into account the stronger suppression due to virtuality implied by
Eq. (\ref{virtqcdg}). We do this by considering the diagram where $\gamma$
splits into a $q \bar q$ pair and then the $q (\bar q)$ of virtuality $q_1^2$
emits a gluon of virtuality $q_2^2>q_1^2$, in the spirit of the backward
showering algorithm. This gives:
\be
\fggam (x,\qsq,\psq)  \sim \int_{\psq}^{\qsq} {dq_1^2 \over q_1^2}
\int_{q_1^2}^{\qsq} {dq_2^2 \over q_2^2}\;\; \alpha_s .
\label{backward}
\ee
The different ans\"atze  now depend on the choice of scale of $\alpha_s$ which
is constrained by the fact that in the limits of $\psq \to 0$ and $\psq \to
\qsq$ the result must give the behaviour implied by Eq. (\ref{virtqcdg}). This
rules out $q_1^2$ as the choice of this scale, but both \qsq\ and
$q_2^2$ as the choice of the scale give acceptable behaviour of
$\fggam (x,\qsq,\psq)$. This gives rise to two different ans\"atze similar to
Eq. (\ref{qansatz1}). The simplest is the former choice and in this case we get
\beq
\fggam^{(1a)} (x,\qsq,\psq) & = \glgam ,\;\;\;\;\;\;\;\; \psq \leq \pcsq
\nonumber \\
& = c_G(x,\qsq)\; {\ln^2 ( \qsq / \psq) \over \ln(\qsq / \Lambda^2)},
\;\;\; \psq \geq \pcsq ,
\label{gansatz1a}
\eeq
Again the continuity of the ansatz at $\psq = \pcsq$ gives us an equation for
$c_g(x,\qsq)$ similar to Eq. (\ref{cq}). Here again, for the ansatz of
Eq. (\ref{gansatz1a}) one can either  choose $P_c$ to be a free parameter
or determine it by requiring that
\be
c_g(x,\qsq)  = c_g^{QPM}(x) =  {\alpha \over \pi} {6 \over 33 -2N_f}
                               \sum_{q,\bar q} e^2_q
\biggl[ {4 \over 3} \biggl( {1\over x} - x^2 \biggr) + 1 -x + 2(1+x)\ln x
\biggr].
\label{cg}
\ee
Here we have to remember that
the dependence on \qsq\ and \psq\ has already been factored out in
Eq. (\ref{gansatz1a}). If we take $q^2_2$ to be the scale of $\alpha_s$ in
Eq. (\ref{backward}), we get
\beq
\fggam^{(1b)} (x,\qsq,\psq) & = \glgam , \;\;\;\; \psq \leq \pcsq \nonumber \\
& = c_G(x,\qsq) \biggl[ \ln {\qsq \over \psq}  - \ln {\psq \over \Lambda^2}
\ln \biggl({\ln (\qsq/\psq) \over \ln(\qsq / \Lambda^2)} \biggr) \biggr],
\;\; \psq \geq \pcsq
\label{gansatz1b}
\eeq
Again both these ans\"atze suffer from the discontinuity in $\ln \psq$; as
before this problem can be solved by writing an ansatz that smoothly
interpolates between the entire \psq\ region:
\be
\fggam^{(2)} (x,\qsq,\psq) = \glgam {\ln^2 {\qsq + \psq \over \psq + \pcsq}
\over \ln^2 \left(1 + {\qsq \over \pcsq}\right)} .
\label{gansatz2}
\ee

As was the case with Eqs. (\ref{qansatz1}) and (\ref{qansatz2}),
substituting Eqs. (\ref{gansatz1a}), (\ref{gansatz1b}) and (\ref{gansatz2}),
in Eq. (\ref{virtie})  we can write the gluon density in the electron again as
a
single integral
\be
\fge (y) = \int_y^1 {dx \over x} \ftilde \left( {y \over x} \right) \qgam \;
{\cal H}_g(\qsq,\psq,\pcsq,\psqmax,\psqmin),
\label{effege}
\ee
where  the function ${\cal H}_g$  can be computed analytically as before
\cite{10}.
Apart from the QPM version of the ansatz (1a) of Eq. (\ref{gansatz1a}), $P_c$
is a free parameter; \psqmax, \psqmin\ depend on the momentum fraction $x$ of
the photon carried by the electron, \qsq, as well as the (anti--)tagging
conditions.

Now  we are ready to give some numerical examples of the suppresions
of the photonic parton densities at TRISTAN energies for
the various ans\"atze given above.
Fig. \ref{fig1} shows the suppression expected for the $u-$quark density in a
no--tag situation for TRISTAN energy for $\qsq = 10\ {\rm GeV}^2$ (which is
the scale relevant for jet production in these experiments), for the GRV
\cite{11} parametrisation of \qgam\ and \glgam . `No--tag' means that
$P^2_{\rm{max, kin}} =  s(1-z)$. The dotted line shows the suppression, w.r.t.
the full unconstrained density given by Eq. (\ref{iel}) using Eq.
(\ref{epa1b}),
from requiring $P^2 < Q^2$ as implied by Eq. (\ref{finlim}). We see that this
condition already suppresses the parton flux by about 20 \%  over most of the
$x$ range. At large values of $x$, $P^2_{\rm {max, kin}}$ itself is small and
hence requiring $P^2 < Q^2$ does not cause further suppression. The short and
long dashed curves show the ansatz of Eq. (\ref{qansatz1}) with fixed
$\pcsq=0.3$ GeV$^2$ and with \pcsq\ estimated from QPM, respectively. The
solid line shows the case of fixed $\pcsq = 0.5\ {\rm GeV}^2 $ for the
smoothed ansatz of Eq. (\ref{qansatz2}). We see that all the various ans\"atze
predict a similar further suppression by about 10\%. The fact that the QPM
case and fixed \pcsq\ case give similar results is  encouraging. The
suppression is reasonably independent of the shape of \qgam\ and hence
is similar for the various parametrisations of \qgam. This suppression will
become more severe with increasing \qsq, since in that case a larger fraction
of the integral in Eq. (\ref{virtie}) will come from the region $\psq > \pcsq$
where the densities \fqgam\ are suppressed.  For the gluons, in a no--tag
situation, one finds on the average marginally less suppression, as compared
to the quark case, coming solely from the dynamical constraint $\psq < \qsq$,
but considerably higher additional suppression due to virtuality effects, of
about 12-15\%.

In Fig. \ref{fig2} we show the suppression expected for \fge\ in an anti--tag
situation with anti--tagging angle of $3.2^\circ$ (for $z < 0.75$,
as used by the TOPAZ collaboration \cite{1}). Here we have used the DG
\cite{12}
parametrisation of the photonic densities. We again show the ratio with the
completely unconstrained densities. The dotted line shows effect
of the dynamical constraint $\psq < \qsq$.  The different dashed, solid and
dash--dotted lines show that the  effects of virtuality cause a further
suppression of as much as 10\%, even in the small $x-$region,
which contributes most to the cross--sections. For the quark case, in the
anti--tag situation, the additional suppression coming from virtuality
effects is only about 5\%.
It is clear that the suppression of double--resolved processes will be
stronger as \fggam\ and \fqgam\ are involved twice. In Fig. \ref{fig3} we see
indeed that the effect of virtuality on the theoretical predictions, for the
rather large anti--tagging angle used by AMY \cite{1}, is considerable and
therefore has to be included in the predictions.

Let us note in passing that for most HERA measurements of photoproduction
cross--sections reported so far, these effects are completely irrelevant as
they use a cut $\psq < 0.1\ {\rm GeV}^2$. However, HERA experiments will soon
have a small angle $e$ tagger which will give events with $ 0.1 < \psq < 1$
${\rm  GeV}^2$. In this situation, at least for our ans\"atze with fixed $P_c$
the suppression of parton densities is completely independent of $x$ as
\psqmax\ and \psqmin\ are fixed, and further this suppression decreases with
\qsq. Taking the ratio of jet--events when the $e$ is tagged in the small
angle tagger to the ones which are tagged in the forward tagger, correcting
for the known difference in the photon fluxes in the two cases, one should be
able to see directly the suppression of the cross--sections due to the
virtuality of the photons \cite{10}.

\vspace*{0.5cm}
\noindent
{\bf Acknowledgements:}
The work of M.D. was supported in part by the U.S. Department of Energy under
grant No. DE-FG02-95ER40896, by the Wisconsin Research Committee with funds
granted by the Wisconsin Alumni Research Foundation, as well as by a grant
from the Deutsche Forschungsgemeinschaft under the Heisenberg program. R.M.G.
wishes to acknowledge a research grant no. 03(0745)/94/EMR-II
from the Council for Scientific and Industrial Research.

\end{document}